%% file: paper.tex
\documentclass[orivec]{llncs}
\usepackage{url,wrapfig,array}
\usepackage[hide]{ed}
\usepackage{amsmath,amssymb} 
\usepackage{tikz}
\usetikzlibrary{shapes}
\usetikzlibrary{arrows}
\usepackage{listings}
\lstMakeShortInline[basicstyle=\tt]|
\usepackage{times}
\usepackage[final,today,eso-foot]{svninfo}

\usepackage{basics}
\usepackage[llncs]{theorems} 
\usepackage{local}
\usepackage[a4paper, bookmarks, bookmarksopen,bookmarksnumbered,
linkcolor=blue, citecolor=blue, urlcolor=blue, colorlinks=true, breaklinks]{hyperref}

\begin{document}
\svnInfo $Id: paper.tex 19570 2010-04-16 12:48:25Z frabe $
\svnKeyword $HeadURL: https://svn.kwarc.info/repos/kwarc/rabe/Papers/omdoc-spec/MKM_10/paper.tex $

\title{Towards MKM in the Large: Modular Representation and Scalable Software Architecture
\thanks{The final publication of this paper is available at www.springerlink.com.}}
\author{Michael Kohlhase, Florian Rabe, Vyacheslav Zholudev}
\institute{Computer Science, Jacobs University Bremen\\
\email{\{m.kohlhase,f.rabe,v.zholudev\}@jacobs-university.de}}
\maketitle

\begin{abstract} 
  MKM has been defined as the quest for technologies to manage mathematical knowledge. MKM
  ``in the small'' is well-studied, so the real problem is to scale up to large, highly
  interconnected corpora: ``MKM in the large''. We contend that advances in two areas are
  needed to reach this goal. We need representation languages that support incremental
  processing of all primitive MKM operations, and we need software architectures and
  implementations that implement these operations scalably on large knowledge bases. 

  We present instances of both in this paper: the {\mmt} framework for modular theory-graphs
  that integrates meta-logical foundations, which forms the base of the next {\omdoc} version; and {\tntbase}, a versioned storage system for XML-based
  document formats. {\tntbase} becomes an {\mmt} database by instantiating it with special MKM operations for {\mmt}. 
\end{abstract}

\section{Introduction}\label{sec:mmttnt:intro}
\input{intro}

\section{A Scalable Representation Language}\label{sec:mmttnt:mmt}
\input{mmt}

\section{A Scalable Implementation}\label{sec:mmttnt:flomdoc}
\input{flomdoc}

\section{A Scalable Database}\label{sec:mmttnt:tnt}
\input{tnt}

\section{Conclusion and Future Work}\label{sec:mmttnt:conclusion}
\input{conc}

\bibliographystyle{plain}
\bibliography{rabe,institutions,systems,pub_rabe,kwarc}

\end{document}

%% file: intro.tex
\svnInfo $Id: intro.tex 19570 2010-04-16 12:48:25Z frabe $
\svnKeyword $HeadURL: https://svn.kwarc.info/repos/kwarc/rabe/Papers/omdoc-spec/MKM_10/intro.tex $

\cite{Farmer:mkm05} defines the objective of MKM to be {\emph{to develop new and better
    ways of managing mathematical knowledge using sophisticated software tools}} and later
states the ``Grand Challenge of MKM'' as {\emph{a universal digital mathematics library
    (UDML)}}, which is indeed a grand challenge, as it envisions that the UDML
{\emph{would continuously grow and in time would contain essentially all mathematical
    knowledge}}, which is estimated to be well in excess of $10^{7}$ published pages.
\footnote{For instance, Zentralblatt Math contains 2.4 million abstracts of articles form
  mathematical journals in the last 100 years.}  All current efforts towards comprehensive
machine-organizable libraries of mathematics are at least three orders of magnitude
smaller than the UDML envisioned by Farmer in 2004: Formal libraries like those of Mizar
(\cite{mizar}, Isabelle (\cite{isabelle}) or PVS (\cite{pvs}) have ca. $10^{4.x}$
statements (definitions and theorems). Even the semi-formal, commercial Wolfram MathWorld
which hails itself {\emph{the world's most extensive mathematics resource}} only has
$10^{4.1}$ entries. There is anecdotal evidence that already at this size, management
procedures are struggling.

To meet the MKM Grand Challenge will have to develop fundamentally more scalable ways of
dealing with mathematical knowledge, especially since \cite{Farmer:mkm05} goes on to
postulate that the UDML {\emph{would also be continuously reorganized and consolidated as new connections and discoveries were made}}. Clearly this can only be achieved
algorithmically; experience with the libraries cited above already show that manual MKM
does not scale sufficiently. Most of the work in the MKM community has concentrated on what we could call ``MKM in the
small'', i.e. dealing with aspects of MKM that do not explicitly address issues of very
large knowledge collections; these we call ``MKM in the large''.

In this paper we contribute to the MKM Grand Challenge of doing formal ``MKM in the large'' by
analyzing scalability challenges inherent in MKM and propose steps towards solutions based on
our {\mmt} format, which is the basis for the next version of \omdoc.  We justify our
conclusions and recommendations for scalability techniques with concrete case studies we
have undertaken in the last years.  Section~\ref{sec:mmttnt:mmt} tackles scalability
issues pertaining to the representation languages used in the formalization of mathematical
knowledge. Section~\ref{sec:mmttnt:flomdoc} discusses how the modularity features of {\mmt}
can be realized scalably by realizing basic MKM functionality like validation, querying,
and presentation incrementally and carefully evaluating the on-the-fly computation (and
caching) of induced representations. These considerations, which are mainly concerned with
efficient computation ``in memory'' are complemented with a discussion of mass storage,
caching, and indexing in Section~\ref{sec:mmttnt:tnt}, which addresses scalability issues
in large collections of mathematical knowledge. Section~\ref{sec:mmttnt:conclusion}
concludes the paper and addresses avenues of further research.


%% file: mmt.tex
\svnInfo $Id: mmt.tex 19570 2010-04-16 12:48:25Z frabe $
\svnKeyword $HeadURL: https://svn.kwarc.info/repos/kwarc/rabe/Papers/omdoc-spec/MKM_10/mmt.tex $
 
Our representation language {\mmt} was introduced in \cite{rabe:thesis:08}. It arises from three central design goals. Firstly, it should provide an expressive but simple \defemph{module system} as modularity is a necessary requirement for scalability. As usual in language design, the goals of simplicity and expressivity form a difficult trade-off that must be solved by identifying the right primitive module constructs. Secondly, scalability across semantic domains requires \defemph{foundation-independence} in the sense that {\mmt} does not commit to any particular foundation (such as Zermelo-Fraenkel set theory or Church's higher-order logic). Providing a good trade-off between this level of generality and the ability to give a rigorous semantics is a unique feature of {\mmt}. Finally, scalability across implementation domains requires \defemph{standards-compliance}, and while using XML and {\openmath} is essentially orthogonal to the language design, the use of URIs as identifiers is not as it imposes subtle constraints that can be very hard to meet a posteriori.

{\mmt} represents logical knowledge on three levels: On the \defemph{module level}, {\mmt} builds on modular algebraic specification languages for logical knowledge such as OBJ \cite{obj}, ASL \cite{asl}, development graphs \cite{devgraphs}, and CASL \cite{caslmanual}. In particular, {\mmt} uses theories and theory morphism as the primitive modular concepts. Contrary to them, {\mmt} only imposes very lightweight assumptions on the underlying language. This leads to a very simple generic module system that subsumes almost all aspects of the syntax and semantics of specific module systems such as PVS \cite{pvs}, Isabelle \cite{isabelle}, or Coq \cite{coq}.

On the \defemph{symbol level}, {\mmt} is a simple declarative language that uses named symbol declarations where symbols may or may not have a type or a definiens. By experimental evidence, this subsumes virtually all declarative languages. In particular, {\mmt} uses the Curry-Howard correspondence \cite{curry,howard} to represent axioms and theorem as constants, and proofs as terms. Sets of symbol declarations yield theories and correspond to {\openmath} content dictionaries.

On the \defemph{object level}, {\mmt} uses the formal grammar of {\openmath} \cite{openmath} to represent mathematical objects without committing to a specific formal foundation. The semantics of objects is given proof theoretically using judgments for typing and equality between objects. {\mmt} is parametric in these judgments, and the precise choice is relegated to a \defemph{foundation}.
 
\subsection{Module System}\label{sec:mmttnt:modules}

Sophisticated mathematical reasoning usually involves several related but different mathematical contexts, and it is desirable to exploit these relationships by moving theorems between contexts. It is well-known that modular design can reduce space to an extent that is exponential in the depth of the reuse relation between the modules, and this applies in particular to the large theory hierarchies employed in mathematics and computer science.

The first applications of this technique in mathematics are found in the works by Bourbaki (\cite{bourbakisets,bourbakialgebra}), which tried to prove every theorem in the context with the smallest possible set of axioms. {\mmt} follows the ``little theories approach'' proposed in \cite{littletheories}, in which separate contexts are represented by separate \defemph{theories}, and structural relationships between contexts are represented as \defemph{theory morphisms}, which serve as conduits for passing information (e.g., definitions and theorems) between theories (see~\cite{farmerintertheory}). This yields \defemph{theory graphs} where the nodes are theories and the paths are theory morphisms.

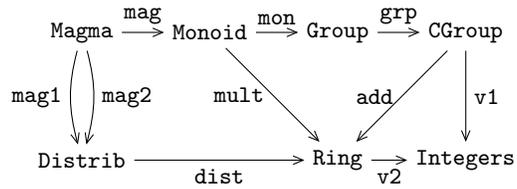
\begin{wrapfigure}{r}{7cm}
\begin{center}\vspace*{-4em}\footnotesize
\begin{tikzpicture}[scale=.85]
\node (M)  at (0,0) {$\cn{Magma}$};
\node (Mo) at (2,0) {$\cn{Monoid}$};
\node (G)  at (4,0) {$\cn{Group}$};
\node (GA) at (6,0) {$\cn{CGroup}$};
\node (R)  at (4,-2){$\cn{Ring}$};
\node (D)  at (0,-2){$\cn{Distrib}$};
\node (I)  at (6,-2){$\cn{Integers}$};
\draw[-\arrowtip](M)  --node[above] {$\cn{mag}$} (Mo);
\draw[-\arrowtip](Mo) --node[above] {$\cn{mon}$} (G);
\draw[-\arrowtip](G)  --node[above] {$\cn{grp}$} (GA);
\draw[-\arrowtip](GA) --node[left] {$\cn{add}$} (R);
\draw[-\arrowtip](Mo) --node[left] {$\cn{mult}$} (R);
\draw[-\arrowtip](M) to[out=-105,in=105] node[left] {$\cn{mag1}$} (D);
\draw[-\arrowtip](M) to[out=-75,in=75] node[right] {$\cn{mag2}$}(D);
\draw[-\arrowtip](D)  --node[below] {$\cn{dist}$} (R);
\draw[-\arrowtip](GA)  --node[right] {$\cn{v1}$} (I);
\draw[-\arrowtip](R)  --node[below] {$\cn{v2}$} (I);
\end{tikzpicture}\vspace*{-1em}
\caption{Algebraic Hierarchy}\label{fig:mmttnt:algebra}\vspace*{-2em}
\end{center}
\end{wrapfigure}

\begin{example}[Algebra]\label{ex:mmttnt:algebra}
For example, consider the theory graph in Fig.~\ref{fig:mmttnt:algebra} for a portion of algebra, which was formalized in {\mmt} in \cite{DHK:algebra:09}. It defines the theory of magmas (A magma has a binary operation without axioms.) and extends it successively to monoids, groups, and commutative groups. Then the theory of rings is formed by importing from both $\cn{CGroup}$ (for the additive operation) and $\cn{Monoid}$ (for the multiplicative operation).

A crucial property here is that the imports are named, e.g., $\cn{Monoid}$ imports from $\cn{Magma}$ via an import named $\cn{mag}$. While redundant in some cases, it is essential in $\cn{Ring}$ where we have to distinguish two theory morphisms from $\cn{Monoid}$ to $\cn{Ring}$: The composition $\cnpath{add,grp,mon}$ for the additive monoid and $\cnpath{mult}$ for the multiplicative monoid.

The import names are used to form qualified names for the imported symbols. For example,
if $*$ is the name of the binary operation in $\cn{Magma}$, then
$\cnpath{add,grp,mon,mag,*}$ denotes addition and $\cnpath{mult,mag,*}$ multiplication in $\cn{Ring}$. Of
course, {\mmt} supports the use of abbreviations instead of qualified names, but it is a crucial prerequisite for scalability to make qualified names the default: Without named imports, every time we add a new name in
$\cn{Magma}$ (e.g, for an abbreviation or a theorem), we would have to add corresponding
renamings in $\cn{Ring}$ to avoid name clashes.

Another reason to use named imports is that we can use them to instantiate imports with
theory morphisms. In our example, distributivity is stated separately as a theory that
imports two magmas. Let us assume, the left distributivity axiom is stated as
\[\forall x,y,z.
   x\;\cnpath{mag1,*}\;(y\;\cnpath{mag2,*}\;z)=(x\;\cnpath{mag1,*}\;y)\;\cnpath{mag2,*}\;(x\;\cnpath{mag1,*}\;z)
\]
Then the import $dist$ from $\cn{Distrib}$ to $\cn{Ring}$ will carry the instantiations
$\maps{\cn{mag1}}{\cnpath{mult,mag}}$ and
$\maps{\cn{mag2}}{\cnpath{\cnpath{add,grp,mon,mag}}}$.

In other module systems such as SML, such instantiations are called (asymmetric) sharing declarations.
In terms of theory morphism, their semantics is a commutative diagram, i.e., an equality between
two morphisms such as $\cnpath{dist,mag1}=\cnpath{mult,mag}:\cn{Magma}\arr\cn{Ring}$. This
provides {\mmt} users and systems with a module level invariant for the efficient structuring of
large theory graphs.

Besides imports, which induce theory morphisms into the containing theory, there is a second kind of edge in the theory graph: Views are explicit theory morphisms that represent translations between two
theories. For example, the node on the right side of the graph represents a theory for the
integers, say declaring the constants $0$, $+$, $-$, $1$, and $\cdot$. The fact that the
integers are a commutative group is represented by the view $\cn{v1}$: If we assume that
$\cn{Monoid}$ declares a constant $e$ for the unit element and $\cn{Group}$ a constant
$\cn{inv}$ for the inverse element, then $\cn{v1}$ carries the instantiations
$\maps{\cnpath{grp,mon,mag,*}}{+}$, $\maps{\cnpath{grp,mon,e}}{1}$, and
$\maps{\cnpath{grp,inv}}{-}$. Furthermore, every axiom declared or imported in
$\cn{CGroup}$ is mapped to a proof of the corresponding property of the integers.

The view $\cn{v2}$ extends $\cn{v1}$ with corresponding instantiations for multiplication. {\mmt} permits modular views as well: When defining $\cn{v2}$, we can import all instantiations of $\cn{v1}$ using $\maps{\cn{add}}{\cn{v1}}$. As above, the semantics of such an instantiation is a commutative diagram, namely $\ö{\cn{add}}{\cn{v2}}=\cn{v1}$ as intended.
\end{example}

The major advantage of modular design is that every declaration --- abbreviations, theorems, notations etc. --- effects \defemph{induced declarations} in the importing theories. A disadvantage is that declarations may not always be located easily, e.g., the addition in a ring is declared in a theory that is four imports away. {\mmt} finds a compromise here: Through qualified names, all induced declarations are addressable and locatable. The process of removing the modularity by adding all these induced declarations to all theories is called \defemph{flattening}.

\begin{experiment}
The formalization in \cite{DHK:algebra:09} uses the Twelf module system (\cite{RS:twelfmod:09}), which is a special case of {\mmt}. Twelf automatically computes the flattened theory graph. The modular theory graph including all axioms and proofs can be written in 180 lines of Twelf code. The flattened graph is computed in less than half a second and requires more than 1800 lines.

The same case study defines two theories for lattices, one based on orderings and one
based on algebra, and gives mutually inverse views to prove the equivalence of the two
theories. Both definitions are modular: Algebraic lattices arise by importing twice from
the theory of semi-lattices; order-based lattices arise by importing the infimum operation
twice, once for the ordering and once for its dual. Consequently, the views can be given
modularly as well, which is particularly important because views must map axioms to
expensive-to-find proofs. These additional declarations take 310 lines of Twelf in modular
and 3500 lines in flattened form.  

These numbers already show the value of modularity in representation already in very small
formalizations. If this is lacking, later steps will face severe scalability problems from
blow-up in representation. Here, the named imports of {\mmt} were the crucial innovation to
strengthen modularity.
\end{experiment}

\subsection{Foundation-Independence}

Mathematical knowledge is described using very different foundations, and the most common foundations can be grouped into set theory and type theory. Within each group there are numerous variants, e.g., Zermelo-Fraenkel or G\"odel-Bernays set theory, or set theories with or without the axiom of choice. Therefore, a single representation language can only be adequate if it is foundation-independent.

{\openmath} and {\omdoc} achieve this by concentrating on structural issues and leaving lexical ones to an external definition mechanism like content dictionaries or theories. In particular, this allows us to operate without choosing a particular foundational logical system, as we can just supply content dictionaries for the symbols in the particular logic. Thus, logics and in the same way foundations become theories, and we speak of the {\defemph{logics-as-theories}} approach.

But conceptually, it is helpful to distinguish levels here. To state a property in the theory $\cn{CGroup}$ like commutativity of the operation $\circ:=\cnpath{grp,mon,mag,*}$ as $\forall a,b.a\circ b=b\circ a$, we use symbols $\forall$ and $=$ from first-order logic together with $\circ$ from $\cn{CGroup}$. Even though it is structurally possible to build algebraic theories by simply importing first-order logic, this would fail to describe the meta-relationship between the theories. But this relation is crucial, e.g., when interpreting $\cn{CGroup}$ in the integers, the symbols of the meta-language are not interpreted because a fixed interpretation is given in the context.

To understand this example better, we use the $M/T$ notation for meta-languages. $M/T$ refers to working in the object language $T$, which is defined within the meta-language $M$. For example, most of mathematics is carried out in $FOL/ZF$, i.e., first-order logic is the meta-language, in which set theory is defined. $FOL$ itself might be defined in a logical framework such as $LF$, and within $ZF$, we can define the language of natural numbers, which yields $LF/FOL/ZF/Nat$. For algebra, we obtain, e.g., $FOL/\cn{Magma}$. {\mmt} makes this meta-relation explicit: Every theory $T$ may point to another theory $M$ as its meta-theory. We can write this as $MMT/(M/T)$.

\begin{wrapfigure}{r}{6cm}\vspace*{-2em}
\fbox{
\begin{tikzpicture}
\node (A) at (0,3)  {$\cn{LF}$};
\node (A') at (2,3)  {$\cn{Isabelle}$};
\node (C) at (-1,1.5)   {$\cn{FOL}$};
\node (C') at (1,1.5) {$\cn{HOL}$};
\node (E) at (-2,0) {$\cn{Monoid}$};
\node (E') at (0,0)  {$\cn{Ring}$};
\draw(A) --node[left] {meta} (C);
\draw(A) --node[right] {meta} (C');
\draw(C) --node[left] {meta} (E);
\draw(C) --node[right] {meta} (E');
\draw[-\arrowtip](A) --node[above] {$m$} (A');
\draw[-\arrowtip](C) --node[above] {$m'$} (C');
\draw[-\arrowtip](E) --node[above] {$\cn{mult}$} (E');
\end{tikzpicture}
}\vspace*{-.5em}
\caption{Meta-Theories}\label{fig:mmt:intro:metatheory}\vspace*{-1em}
\end{wrapfigure}
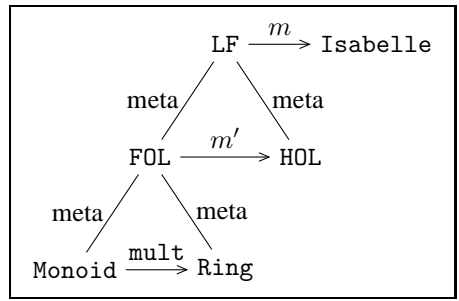

In Fig.~\ref{fig:mmt:intro:metatheory}, the algebra example is extended by adding meta-theories. The theory $\cn{FOL}$ for first-order logic is the meta-theory for all algebraic theories, and the theory $\cn{LF}$ for the logical framework LF is the meta-theory of $\cn{FOL}$ and of the theory $\cn{HOL}$ for higher-order logic.

Now the crucial advantage of the logics-as-theories approach is that on all three levels the same module system can be used: For example, the views $m$ and $m'$ indicate possible translations on the levels of logical frameworks and logics, respectively. Similarly, logics and foundations can be built modularly. Thus, we can use imports to represent inheritance at the level of logical foundations and views to represent formal translations between them. Just like in the little theories approach, we can prove meta-logical results in the simplest foundation that is expressive enough and then use views to move results between foundations.

\begin{example}[Little Logics and Little Foundations]\label{ex:mmttnt:foundations}
In \cite{HR:folsound:09}, we formalize the syntax, proof theory, and model theory and prove the soundness of first-order logic in {\mmt}. Using the module system, we can treat all connectives and quantifiers separately. Thus, we can reuse these fragments to define other logics, and in \cite{project:latin} we do that, e.g., for sorted first-order logic and modal logic. 

For the definition of the model theory, we need to formalize set theory in {\mmt}, which is a significant investment, and even then doing proofs in set theory --- as needed for the soundness proof --- is tedious. Therefore, in \cite{HR:folsound:10}, we develop the set theoretical foundation itself modularly. We define a typed higher-order logic $\cn{HOL}$ first, which is expressive enough for many applications such as the above soundness proof. Then a view from $\cn{HOL}$ to $\cn{ZF}$ proves that $\cn{ZF}$ is a refinement of $\cn{HOL}$ and completes the proof of the soundness of FOL relative to models defined in $\cn{ZF}$.
\end{example}

\begin{experiment}
Ex.~\ref{ex:mmttnt:foundations} already showed that it is feasible to represent foundations and relations between foundations in {\mmt}. Being able to this is a qualitative aspect of cross-domain scalability. In another case study, we represented $LF/Isabelle$ and $LF/Isabelle/HOL$ (\cite{isabelle,isabellehol}) as well as a translation from them into $LF/FOL/ZFC$ (see \cite{project:latin}).
\end{experiment}

To our knowledge, {\mmt} is the only declarative formalism in which comparable foundation or logic translations have been conducted. In Hets (\cite{hets}) a number of logic translations are implemented in Haskell. Twelf and Delphin provide logic and functional programming languages, respectively, on top of LF (\cite{twelf,delphin}), which have been used to formalize the HOL-Nuprl translation (\cite{hol_nuprl}).

\subsection{Symbol Identifiers ``in the Large''}

In mathematical languages, we need to be able to refer to (i.e., identify) content objects in order to state the semantic relations. It was a somewhat surprising realization in the design of {\mmt} that to understand the symbol identifiers is almost as difficult as to understand the whole module system. Theories are containers for symbol declarations, and relations between theories define the available symbols in any given theory. Since every available symbol should have a canonical identifier, the syntax of identifiers is inherently connected to the possible relations between theories.

In principle, there are two ways to identify content object: \defemph{by location} (relative to a particular document or file) and \defemph{by context} (relative to a mathematical theory). The first one essentially makes use of the organizational structure of files and file systems, and the second makes use of mathematical structuring principles supplied by the representation format.

As a general rule, it is preferable to use identification by context as the distribution of
knowledge over file systems is usually a secondary consideration. Then the mapping between theory identifiers and physical theory locations can be deferred to an extralinguistic catalog. Resource identification by context should still be compatible with the URI-based approach that mediates most resource transport over the internet. This is common practice in scalable programming languages such as Java where package identifiers are URIs and classes are located using the \texttt{classpath}.

For logical and mathematical knowledge, the {\openmath} 2 standard (\cite{openmath}) and the current {\omdoc} version 1.2 define URIs for symbols. A symbol is identified by the symbol name and content dictionary, which in turn is identified by the CD name and the CD base, i.e., the URI where the CD is located. From these constituents, symbol URIs are formed using URI fragments (the part after the \# delimiter). However, {\openmath} imposes a one-CD-one-file restriction, which is too restrictive in general. While {\omdoc}1.2 permits multiple theories per file, it requires file-unique identifiers for all symbols. In both cases, the use of URI fragments, which are resolved only on the client, forces clients to retrieve the complete file even if only a single symbol is needed.

Furthermore, many module systems have features that impede or complicate the formation of canonical symbol URIs. Such features include unnamed imports, unnamed axioms, overloading, opening of modules, or shadowing of symbol names. Typically, this leads to a non-trivial correspondence between user-visible and application-internal identifiers. But this impedes or complicates cross-application scalability where all applications (ranging from, e.g., a Javascript GUI to a database backend) must understand the same identifiers.
  
{\mmt} avoids the above pitfalls and introduces a simple yet expressive web-scalable syntax for symbol identifiers. An {\mmt}-URI is of the form $doc?mod?sym$ where
\vspace{-.5em}
\begin{itemize}
\item $doc$ is a URI without query or fragment, e.g.,
  \url{http://cds.omdoc.org/math/algebra/algegra1.omdoc} which
  identifies (but not necessarily locates) an {\mmt} document,
\item $mod$ is a $/$-separated sequence of local names that gives the path to a nested
  theory in the above document, e.g., $\cn{Ring}$,
\item $sym$ is a $/$-separated sequence $imp_1/\ldots/imp_n/con$ of local names such that
  $imp_i$ is an import and $con$ a symbol name, e.g., $\cnpath{mult,mon,*}$,
\item a local name is of the form $pchar^+$ where $pchar$ is defined as in RFC
  3986~\cite{uri}, which --- possibly via \%-encoding --- permits almost all Unicode
  characters.
\end{itemize}

In our running example, the canonical URI of multiplication in a ring is
\url{http://cds.omdoc.org/math/algebra/algegra1.omdoc?Ring?mult/mon/*}.\\  Note that the use
of two \url{?} characters in a URI is unusual outside of {\mmt}, but legal w.r.t. RFC 3986.  Of
course, {\mmt} also defines relative URIs that are resolved against the URI of the
containing declaration. The most important case is when $doc$ is empty. Then the
resolution proceeds as in RFC 3986, e.g., $?mod'?sym'$ resolves to $doc?mod'?sym'$
relative to $doc?mod?sym$ (Note that this differs from RFC 2396.). {\mmt} defines some
additional cases that are needed in mathematical practice and go beyond the expressivity
of relative URIs: Relative to $doc?mod?sym$, the resolution of $??sym'$ and $?/mod'?sym'$
yields $doc?mod?sym'$ and $doc?mod/mod'?sym'$, respectively.

\begin{experiment}
URIs are the main data structure needed for cross-application scalability, and our experience shows that they must be implemented by almost every peripheral system, even those that do not implement {\mmt} itself. Already at this point, we had to implement them in SML (\cite{RS:twelfmod:09}), Javascript (\cite{GLR:jobad:09}), XQuery (\cite{ZKR:tntbase:10}), Haskell (for Hets, \cite{hets}), and Bean Shell (for a jEdit plugin) --- in addition to the Scala-based reference API (Sect.~\ref{sec:mmttnt:flomdoc}).

This was only possible because {\mmt}-URIs constitute a well-balanced trade-off between mathematical rigor, feasibility, and URI-compatibility: In particular, due to the use of the two separators $/$ and $?$ (rather than only one), they can be parsed locally, i.e., without access to or understanding of the surrounding {\mmt} document. And they can be dereferenced using standard URI libraries and URI-URL translations. At the same time, they provide canonical names for all symbols that are in scope, including those that are only available through imports.
\end{experiment}


%% file: flomdoc.tex
\svnInfo $Id: flomdoc.tex 19570 2010-04-16 12:48:25Z frabe $
\svnKeyword $HeadURL: https://svn.kwarc.info/repos/kwarc/rabe/Papers/omdoc-spec/MKM_10/flomdoc.tex $
 
As the implementation language for the {\mmt} reference API, we pick Scala, a programming
language designed to be \emph{scala}ble (\cite{scala}). Being functional, Scala permits
elegant code, and based on and fully compatible with Java, it offers cross-application and
web-level scalability.

The {\mmt} API implements the syntax and semantics of {\mmt}. It compiles to a 1 MB Java
archive file that can be readily integrated into applications. Library and documentation
can be obtained from~\cite{project:mmt}.  Two technical aspects are especially notable
from the point of view of scalability. Firstly, all {\mmt} functionality is exposed to
non-Java applications via a scriptable shell and via an HTTP servlet. Secondly, the API
maintains an abstraction layer that separates the backends that physically store {\mmt}
documents (URLs) from the document identifiers (URIs). Thus, it is configurable which
{\mmt} documents are located, e.g., in a remote database or on the local file
system. In the following section we describe some of the advanced features.

\subsection{Validation}

Validation describes the process of checking {\mmt} theory graphs against the {\mmt}
grammar and type system. {\mmt} validation is done in three increasingly strict \defemph{stages}.

The first stage is XML validation against a context-free RelaxNG grammar.
As this only catches errors in the
surface syntax, {\mmt} documents are validated \defemph{structurally} in a second stage. Structural
validity guarantees that all declarations have unique URIs and that all references to
symbols, theories, etc. can be resolved. This is still too lax for mathematics as it lacks
type-checking. But it is exactly the right middle ground between the weak
validation against a context-free grammar and the expensive and complex validation against
a specific type system: On the one hand, it is efficient and foundation-independent, and
on the other hand, it provides an invariant that is sufficient for many MKM services such
as storage, navigation, or search.

Type-checking is foundation-specific, therefore each foundation must provide an {\mmt} plugin that implements the typing and equality judgments.  More precisely, the plugin must provide function that (semi-)decide for two given terms $A$ and $B$ over a theory $T$, the judgments $\vdash_T A=B$ and $\vdash_T A:B$. Given such a plugin, a third validation stage can refine structural validity by additionally validating well-typedness of all declarations. For scalability, it is important that (i) these plugins are stateless as the theory graph is maintained by {\mmt}, and that the (ii) modular
structure is transparent to the plugin. Thus plugin developers only need to provide the
core algorithms for the specific type system, and all MKM issues can be relegated to dedicated implementations.

Context-free validation is well-understood. Moreover, {\mmt} is designed such that foundation-specific validation is obtained from structural validation by using the same inference system with some typing and equality constraints added. This leaves structural validation as the central issue for scalability.

\begin{experiment}
We have implemented structural validation by decomposing an {\mmt} theory graph into a list of atomic declarations. For example, the declaration $T=\{s_1:\tau_1,\;s_2:\tau_2\}$ of a theory $T$ with two typed symbols yields the atomic
declarations $T=\{\}$, $T?s_1:\tau$, and $T?s_2:\tau_2$. This ``unnesting'' of declarations is a special property of the {\mmt} type system that is not usually found in other systems. It is possible because every declaration has a canonical URI and can therefore be taken out of its context.

This is important for scalability as it permits \defemph{incremental} processing. In particular, large {\mmt} documents can be processed as streams of atomic declarations. Furthermore, the semantics of {\mmt} guarantees that the processing order of these streams never matters if the (easily-inferrable) dependencies between declarations are respected. This would even permit parallel processing, another prerequisite for scalability.
\end{experiment}

\subsection{Querying}\label{sec:mmttnt:querying}

Once a theory graph has been read, {\mmt} provides two ways how to access it: {\mmt}-URI dereferencing and querying with respect to a simple ontology.

Firstly, a theory graph always has two forms: the modular form where all nodes are partial
theories whose declarations are computed using imports, and the flattened form where all
imports are replaced with translated copies of the imported declarations. Many implementations of module systems, e.g.,
Isabelle's locales, automatically compute the flat form and do not maintain the modular
form. This can be a threat to scalability as it can induce combinatorial explosion.

{\mmt} maintains only the modular form. However, as every declaration present in the flat form has a canonical URI, the access to the flat form is possible via {\mmt}-URI dereferencing: Dereferencing computes (and caches) the induced declarations present in the flat form. Thus, applications can ignore the modular structure and interact with a modular theory graph as if it were flattened, but the exponentially expensive flattening is performed transparently and incrementally.

Secondly, the API computes the ABox of a theory graph with respect to the {\mmt} ontology. It has {\mmt}-URIs as individuals and 10 types like \texttt{theory} or \texttt{symbol} as unary predicates. 11 binary predicates represent relations between individuals such as \texttt{HasDomain} relating an import to a theory or \texttt{HasOccurrenceOfInType} relating two symbols. These relations are structurally complete: The structure of a theory graph can be recovered from the ABox. The computation time is negligible as it is a byproduct of validation anyway.

The API includes a simple query language for this ontology. It can compute all individuals in the theory graph that are in a certain relation to a given individual. The possible queries include composition, union, transitive closure, and inverse of relations.

The ABox can also be regarded as the result of \defemph{compiling} an {\mmt} theory graph. Many operations on theory graphs only require the ABox: for example the computation of the forward or backward dependency cone of a declaration which are needed to generate self-contained documents and in change management, respectively. This is important for cross-application scalability because applications can parse the ABox very easily.
Moreover, we obtain a notion of \defemph{separate compilation}: ABox-generation only requires structural validity, and the latter can be implemented if only the ABoxes of the referenced files are known.

\begin{experiment}
Since all {\mmt} knowledge items have a globally unique {\mmt}-URI, being able to dereference them is sufficient to obtain complete information about a theory graph. We have implemented a web servlet for {\mmt} that can act as a local proxy for {\mmt}-URIs and as a URI catalog that maps {\mmt}-URIs into (local or remote) URLs. The former means that all {\mmt}-URIs are resolved locally if possible. The latter means that the {\mmt}-URI of a module can be independent from its physical location. The same servlet can be run remotely, e.g., on the same machine as a mathematical database and configured to retrieve files directly from there or from other servers.

Thus systems can access all their input documents by URI via a local service, which makes all storage issues transparent. (Using presentation, see below, these can even be presented in the system's native syntax.) This solves a central problem in current implementations of formal systems: the restriction to in-memory knowledge. Besides the advantages of distributed storage and caching, a simple example application is that imported theories are automatically included when remote documents are retrieved in order to avoid successive lookups.
\end{experiment}

\subsection{Presentation}\label{sec:mmttnt:presentation}

{\mmt} comes with a declarative language for notations similar to \cite{KMR:presentation:08} that can be used to transform {\mmt} theory graphs into arbitrary output formats. Notations are declared by giving parameters such as fixity and input/output precedence, and snippets for separators and brackets. Notations are not only used for mathematical objects but also for all {\mmt} expressions, e.g. theory declarations and theory graphs.

Two aspects are particularly important for scalability. Firstly, sets of notations (called
\emph{styles}) behave like theories, which are sets of symbols. In particular, styles and
notations have {\mmt}-URIs (and are part of the {\mmt} ontology), and the {\mmt} module
system can be used for inheritance between styles.

Secondly, every {\mmt} expression has a URI $E$, for declarations this is trivial, for most mathematical objects it is the URI of the head symbol. Correspondingly, every notation must give an {\mmt}-URI $N$, and the notation is applicable to an expression if $N$ is a prefix of $E$. More specific notations can inherit from more general ones, e.g., the brackets and separators are usually given only once in the most general notation. This simplifies the authoring and maintenance of styles for large theory graphs significantly.

\begin{experiment}
  In order to present {\mmt} content as, e.g., HTML with embedded presentation {\mathml},
  we need a style with only the 20 generic notations given in
  \url{http://alpha.tntbase.mathweb.org/repos/cds/omdoc/mathml.omdoc}.

\begin{wrapfigure}r{5.6cm}
\begin{minipage}{5.5cm}
\begin{lstlisting}[belowskip=-3em]
<notation for="http://cds.omdoc.org/" 
                 role="constant">
  <element name="mo">
    <attribute name="xref">
      <text value="#"/><id/>
    </attribute>
    <hole><component index="2"/></hole>
  </element>
</notation>
\end{lstlisting}
\end{minipage}
\end{wrapfigure}
 For example, the notation declaration on the right applies to all constants whose
  \texttt{cdbase} starts with \url{http://cds.omdoc.org/} and renders \texttt{OMS}
  elements as \texttt{mo} elements. The latter has an \texttt{xref} attribute that
  links to the parallel markup (which is included by notations at higher levels). The
  content of the \texttt{mo} elements is a ``hole'' that is by default filled with the
  second component, for constants that is the name ($0$ and $1$ are \texttt{cdbase} and
  \texttt{cd}.).

This scales well because we can give notations for specific theories, e.g., by saying that $?Magma?*$ is associative infix and optionally giving a different operator symbol than $*$. We can also add other output formats easily: Our implementation (see \cite{project:latin}) extends the above notation with a \texttt{jobad:href} attribute containing the {\mmt}-URI --- this URI is picked up by our JOBAD Javascript (\cite{GLR:jobad:09}) for hyperlinking.
\end{experiment}


%% file: tnt.tex
\svnInfo $Id: tnt.tex 19570 2010-04-16 12:48:25Z frabe $
\svnKeyword $HeadURL: https://svn.kwarc.info/repos/kwarc/rabe/Papers/omdoc-spec/MKM_10/tnt.tex $

The {\tntbase} system~\cite{tntbase} is a versioned XML-database with a client-server architecture.
It integrates Berkeley DB XML into a Subversion server. DB XML stores HEAD revisions of XML files; non-XML content like PDF, images or {\LaTeX} source files, differences between revisions, directory entry lists and other repository information are retained in a usual SVN back-end storage (Berkeley DB in our case).
Keeping XML documents in DB XML allows accessing files not only via any SVN client but also through the DB XML API that supports efficient querying of XML content via XQuery and (versioned) modification of that content via XQuery Update.

In addition, {\tntbase} provides a plugin architecture for document
format-specific customizations~\cite{ZKR:tntbase:10}. Using {\omdoc}
as concrete syntax for {\mmt} and the {\mmt} API as a {\tntbase}
plugin, we have made {\tntbase} {\mmt}-aware so that data-intensive {\mmt} algorithms can be executed within the database.

The {\tntbase} system and its documentation are available at
\url{http://tntbase.org}. Below we describe some of the features particularly relevant for scalability.

\subsection{Generating Content}

Large scale collaborative authoring of mathematical documents requires \defemph{distributed} and \defemph{versioned} storage. On the language end, {\mmt} supports this by making all identifiers URIs so that {\mmt} documents can be distributed among authors and networks and reference each other. On the database end, {\tntbase} supports this by acting as a versioned {\mmt} database.

In principle, versioning and distribution could also be realized with a plain SVN server. But for mathematics, it is important that the storage backend is aware of at least some aspects of the mathematical semantics. In large scale authoring processes, an important requirement is to guarantee consistency, i.e., it should be possible to reject commits of invalid documents. Therefore, {\tntbase} supports document format-specific \defemph{validation}.

For scalability, it is crucial that validation of interlinked collections of {\mmt} documents is \defemph{incremental}, i.e., only those documents added or changed during a commit are validated. This is a significant effect because the committed documents almost always import modules from other documents that are already in the database, and these should not be revalidated --- especially not if they contain unnecessary modules that introduce further dependencies. 

Therefore, we integrate {\mmt} separate compilation into {\tntbase}. During a commit {\tntbase} validates all committed files structurally by calling the {\mmt} API. After successful validation, the ABox is generated and immediately stored in {\tntbase}. References to previously committed files are not resolved; instead their generated ABox is used for validation. Thus, validation is limited to the committed documents.

\begin{experiment}
In the {\latin} project~\cite{project:latin}, we create an atlas of
logics and logic translations formalized in {\mmt}. At the current
early stage of the project 5 people are actively editing so far about
100 files. These contain about 200 theories and 50 views, which form a
single highly inter-linked {\mmt} theory graph. We use {\tntbase} as the validity-guaranteeing backend storage.

The {\latin} theory graph is highly modular. For example, the document giving the
set-theoretical model theory of first-order logic from
\cite{HR:folsound:10} depends on about 100 other theories. (We counted
them conveniently using an XQuery, see below.) Standalone validation
of this document takes about 15 seconds if needed files are retrieved
from a local file system. Using separate compilation in {\tntbase}, it
is almost instantaneous.
\end{experiment}

In fact, we can configure {\tntbase} so that structural validation is
preceded by RelaxNG validation. This permits the {\mmt} application to
drop inefficient checks for syntax errors. Similarly, structural
validation could be preceded by foundation-specific validation, but
often we do not have a well-understood notion of separate compilation
for specific foundations. But even in that case, we can do better than
naive revalidation. {\mmt} is designed so that it is
foundation-independent which modules a given document depends
on. Thus, we can collect these modules in one document using an
efficient XQuery (see below) and then revalidate only this
document. Moreover, we can use the presentation algorithm from
Sect.~\ref{sec:mmttnt:presentation} to transform the generated
document into the input syntax of a dedicated implementation.

\subsection{Retrieving Content}

While the previous section already showed some of the strength of an {\mmt}-aware {\tntbase}, its true strength lies in retrieving content. As every XML-native database, {\tntbase} supports XQuery but extends the DB XML syntax by a notion of file system path to address path-based collections of documents. Furthermore, it supports indexing to improve performance of queries and the querying of previous revisions. Finally, custom XQuery modules can be integrated into {\tntbase}.

{\mmt}-aware retrieval is achieved through two measures. Firstly, \defemph{ABox caching} means that for every committed file, the {\mmt} ABox is generated and stored in {\tntbase}. The ABox contains only two kinds of declarations --- instances of unary and binary predicates --- and is stored as a simple XML document. The element types in these documents are \defemph{indexed}, which yields efficient global queries.

\begin{example}
An {\mmt} document for the algebra example from Sect.~\ref{sec:mmttnt:modules} is served at
{\footnotesize http://alpha.tntbase.mathweb.org/repos/cds/math/algebra/algebra1.omdoc}. Its ABox is cached at
{\footnotesize http://alpha.tntbase.mathweb.org:8080/tntbase/cds/restful/integration/validation/mmt/content/ math/algebra/algebra1.omdoc}.
\end{example}

Secondly, custom \defemph{XQuery functions} utilize the cached and indexed ABoxes to provide efficient access to frequently needed aggregated documents. These include in particular the forward and backward dependency cones of a module. The backward dependency cone of a module $M$ is the minimal set of modules needed to make $M$ well-formed. Dually, the forward cone contains all modules that need $M$. If it were not for the indexed ABoxes, the latter would be highly expensive to compute: linear in the size of the database.

\begin{experiment}
The {\mmt} presentation algorithm described in Sect.~\ref{sec:mmttnt:presentation} takes a set of notations as input. However, additional notations may be given in imported theories, typically format-independent notations such as the one making $?\cn{Magma}?*$ infix. Therefore, when an {\mmt} expression is rendered, all imported theories must be traversed for the sole reason of obtaining all notations.

Without {\mmt} awareness in {\tntbase}, this would require multiple
successive queries which is particularly harmful when presentation is
executed locally while the imported theories are stored remotely. But even when all theories are available on a local disk, these successive calls already take 1.5 seconds for the above algebra document. (Once the notations are retrieved, the presentation itself is instantaneous.)

In {\mmt}-aware {\tntbase}, we can retrieve all notations in the
backward dependency closure of the presented expression with a single XQuery. ABox-indexing made this instantaneous up to network lag.
\end{experiment}
\smallskip

{\tntbase} does not only permit the efficient retrieval of such generated documents, but it also permits to commit edited versions of them. We call these \defemph{virtual documents} in \cite{ZKR:tntbase:10}. These are essentially ``XML database views'' analogous to views in relational databases. They are editable, and {\tntbase} transparently patches the differences into the original files in the underlying versioning system.

\begin{experiment}
While manual refactoring of large theory graphs is as difficult as refactoring large software, there is virtually no tool support for it. For {\mmt}, we obtain a simple renaming tool using a virtual document for the one-step (i.e., non-transitive) forward dependency cone of a theory $T$ (see \cite{ZKR:tntbase:10} for an example). That contains all references to $T$ so that $T$ can be renamed and all references modified in one go.
\end{experiment}



%% file: conc.tex
\svnInfo $Id: conc.tex 19277 2010-03-11 03:44:12Z frabe $
\svnKeyword $HeadURL: https://svn.kwarc.info/repos/kwarc/rabe/Papers/omdoc-spec/MKM_10/conc.tex $

This paper aims to pave the way for MKM ``in the large'' by proposing a theoretical and
technological basis for a ``Universal Digital Mathematics Library'' (UDML) which has been
touted as the grand challenge for MKM. In a nutshell, we conclude that the problem of
scalability has be to addressed on all levels: we need modularity and accessibility of
induced declarations in the representation
format, incrementality and memoization in the implementation of the fundamental
algorithms, and a mass storage solution that supports fragment access and indexing. We
have developed prototypical implementations and tested them on a variety of case studies.

The next step will be to integrate the parts and assemble a UDML installation with
these. We plan to use the next generation of the \omdoc format, which will integrate the
MMT infrastructure described in this paper as an interoperability layer;
see~\cite{KRS:integration:10} for a discussion of the issues involved.  In the last years,
we have developed \omdoc translation facilities for various fully formal theorem proving
systems and their libraries. In the LATIN project~\cite{project:latin}, we are already
developing a graph of concrete ``logics-as-theories'' to make the underlying logics
interoperable.

